\documentclass{iopjournal}
\usepackage{here}
\usepackage{soul}
\begin{document}
\articletype{Paper} 

\title{Teaching light absorption and the Beer-Lambert law using everyday materials: a tomato juice experiment for introductory physics}
\author{Hiroki Wadati$^{1,2,*}$\orcid{0000-0001-5969-8624}}

\affil{$^1$Department of Material Science, Graduate School of Science, University of Hyogo, Ako, Hyogo 678-1297, Japan}

\affil{$^2$Institute of Laser Engineering, Osaka University, Suita, Osaka 565-0871, Japan}

\affil{$^*$Author to whom any correspondence should be addressed.}

\email{wadati@sci.u-hyogo.ac.jp}


\begin{abstract}
I describe a visually engaging experiment to demonstrate the Beer-Lambert law in introductory physics or general science courses. By using diluted tomato juice as a naturally colored absorber and a halogen lamp as a broadband light source, students can explore how light attenuation depends on concentration and wavelength. The activity connects the optical concept of exponential absorption with everyday materials, making it accessible for classrooms with limited resources. Transmission spectra obtained with a compact spectrometer reveal a strong absorption band around 500 nm, corresponding to the green-blue region absorbed by lycopene. Plotting absorbance against concentration allows students to confirm linear behavior at low concentrations and to discuss deviations at higher ones. The experiment emphasizes conceptual understanding of light–matter interaction, quantitative data analysis, and the limitations of ideal laws, providing a memorable learning experience in optics and spectroscopy.
\end{abstract}

\section{Introduction}
The interaction of light with matter provides a powerful and intuitive way for students to explore fundamental physical principles. One of the most straightforward and most instructive examples is the Beer–Lambert law, which describes how the intensity of transmitted light decreases exponentially as it passes through an absorbing medium \cite{1,2}. This relationship between absorbance, concentration, and path length not only underpins spectroscopic analysis but also illustrates the general concept of exponential attenuation that appears across many areas of physics. Observing how a colored liquid absorbs light provides a tangible way for students to connect mathematical models with familiar experiences—such as the fading of color when a solution is diluted. Such demonstrations bridge theory and perception, making the abstract concept of light–matter interaction accessible in introductory physics and general science courses.

Many classroom demonstrations of the Beer–Lambert law employ commercial spectrophotometers or prepared chemical solutions, which can limit accessibility in schools with restricted budgets. To make this fundamental concept more approachable, several educators have introduced alternatives using food dyes, colored beverages, or household materials \cite{3,4}. These experiments successfully engage students by linking quantitative optical measurements with visually striking color changes, helping them grasp how absorbance varies with concentration. Building on these ideas, I present a simple, environmentally safe activity that uses diluted tomato juice as the absorbent medium and a halogen lamp as the light source. Tomato juice contains lycopene \cite{5,6,7}, a naturally occurring pigment with a strong absorption band in the green-blue region of the spectrum, making it an ideal everyday example of selective light absorption. The combination of familiar materials, clear color contrast, and quantitative analysis allows students to experience the Beer–Lambert law directly while appreciating its physical meaning beyond textbook equations. Similar approaches for demonstrating light attenuation have been described in Ref.~\cite{8}, where the authors developed a simple Beer–Lambert law measurement. 
Related educational studies using everyday substances have also been reported, for example, the undergraduate laboratory investigation of olive oil using UV-visible spectroscopy in Ref.~\cite{9}. 

The goal of this study is to provide physics and general science educators with an accessible, visually engaging experiment to introduce the Beer-Lambert law and the concept of light attenuation. By combining everyday materials with quantitative optical measurements, the activity bridges theoretical physics and real-world observation. Students can not only verify the proportional relationship between absorbance and concentration but also explore the conditions under which the law begins to fail, thereby deepening their understanding of measurement limits and experimental uncertainty. The exercise promotes active inquiry, data analysis, and discussion, aligning with inquiry-based learning principles emphasized in modern physics education. Because all materials-tomato juice, tap water, a halogen lamp, and a compact spectrometer-are safe, the activity can be implemented in diverse teaching environments, from high-school demonstrations to undergraduate laboratory courses. Ultimately, this experiment aims to cultivate conceptual understanding and scientific curiosity by showing how a simple optical setup can reveal both the precision and the imperfections of a fundamental physical law.

\section{Experimental Method}
The experiment was designed to be easily implemented in high school and undergraduate laboratory settings using safe materials. A commercially available unsalted tomato juice was used as the absorbing medium, and tap water served as the solvent to prepare a dilution series of seven samples (0, 2.5, 5, 10, 25, 50, and 100 \% v/v). Each solution was poured into a 1 cm path-length plastic cuvette, and the color gradient provided an immediate qualitative illustration of the concentration dependence.

A halogen lamp (Thorlabs QTH10/M, 10 W) served as a broadband light source, and a compact spectrometer (Thorlabs CCS200/M, 200–1000 nm) measured the transmitted light intensity. The lamp and spectrometer were aligned on a simple optical bench with the cuvette placed between them, as shown in Figure 1. A reference spectrum was obtained with water alone, and the transmission spectra of all samples were recorded over 1-100 ms integration times, depending on concentration. The absorbance $A(\lambda)$ was calculated from
\begin{equation}
A(\lambda) = -\log_{10}[I(\lambda)/I_0(\lambda)],
\end{equation}
where $I(\lambda)$ and $I_0(\lambda)$ represent the transmitted and reference intensities, respectively.

\begin{figure}[H]
\centering
\includegraphics[width=0.5\textwidth]{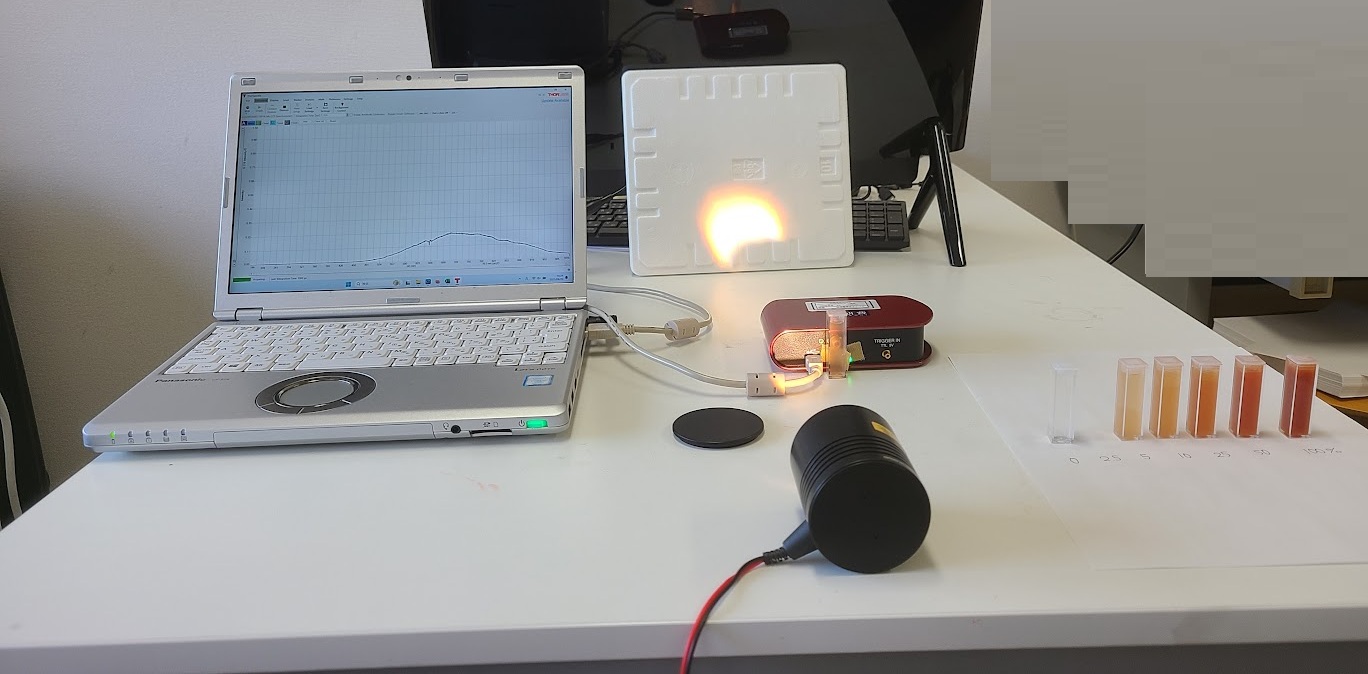}
\caption{Photograph of the experimental setup used in the classroom activity, showing the halogen lamp (Thorlabs QTH10/M), plastic cuvette (MonotaRO 03004285), compact spectrometer (Thorlabs CCS200/M), and the PC for data acquisition. The simple arrangement makes it easy for students to visualize how light passes through the sample.}
\label{fig1}
\end{figure}

In a classroom implementation, students worked in pairs to measure and analyze the data within a 90-minute session. They plotted absorbance versus concentration at selected wavelengths and discussed where the linear behavior predicted by the Beer–Lambert law began to deviate. The activity emphasized teamwork, safe laboratory practice, and critical discussion of measurement errors. Since all equipment is portable and robust, the setup can be reproduced without a dedicated optics laboratory, allowing educators to integrate the exercise into regular physics or general-science courses.

\section{Results and Discussion}
The color gradient in the tomato-juice dilution series immediately demonstrates how concentration affects light absorption, providing students with an intuitive visual introduction to the Beer–Lambert law. Figure 2 shows that as the tomato juice becomes more concentrated, the transmitted light decreases and the color deepens to red. Students can easily connect this everyday observation to the physical concept of selective absorption: lycopene molecules absorb green-blue wavelengths around 500 nm, leaving the transmitted light enriched in red components.

\begin{figure}[H]
\centering
\includegraphics[width=0.5\textwidth]{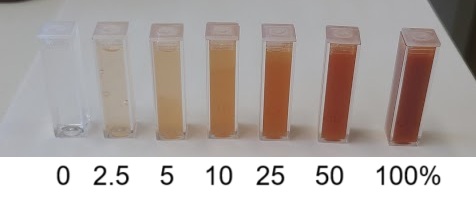}
\caption{Photograph of the seven cuvettes containing the tomato juice dilution series (0, 2.5, 5, 10, 25, 50, and 100\% by volume). The color gradient provides a clear visual representation of concentration differences for students.}
\label{fig2}
\end{figure}

Transmission and absorbance spectra obtained from the seven samples reveal a pronounced absorption band between 480 and 520 nm, characteristic of lycopene, as shown in Figure 3. When students convert the transmission data to absorbance and plot absorbance versus concentration, they can quantitatively explore how the Beer–Lambert law applies. Figure 4(a) presents the absorbance at five representative wavelengths (500, 600, 700, 800, and 900 nm) across the entire concentration range. The plots clearly show an overall increase in absorbance with concentration, confirming the expected proportionality between these quantities.
To highlight this relationship more clearly, Figure 4(b) magnifies the dilute region (0–25 \% v/v). In this range, the data points fall nearly on a straight line, illustrating the linear behavior predicted by the Beer–Lambert law. At higher concentrations, however, the slope begins to deviate from linearity as light scattering, turbidity, and molecular aggregation become significant. Discussing these deviations in class helps students appreciate that physical laws are idealizations valid only within certain limits. Teachers can use this point to introduce real-world measurement issues—such as multiple scattering or detector saturation—and to encourage students to compare the data with their fitted linear models.

\begin{figure}[H]
\centering
\includegraphics[width=0.5\textwidth]{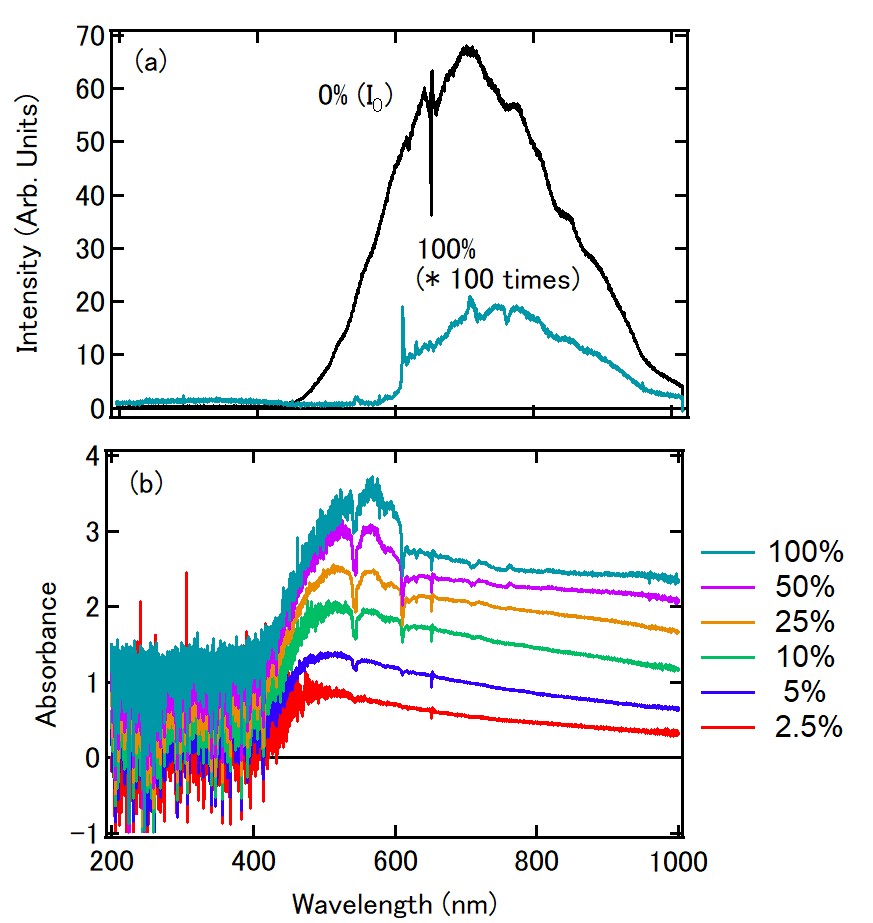}
\caption{(a) Transmission spectra of 0\% (water) and 100\% tomato juice, illustrating the pronounced absorption band around 480-520 nm due to lycopene. (b) Absorbance spectra of all seven tomato juice dilutions, plotted as a function of wavelength. The systematic increase in absorbance with concentration demonstrates the applicability of the Beer-Lambert law across the visible range.}
\label{fig:spectra}
\end{figure}

\begin{figure}[H]
\centering
\includegraphics[width=0.8\textwidth]{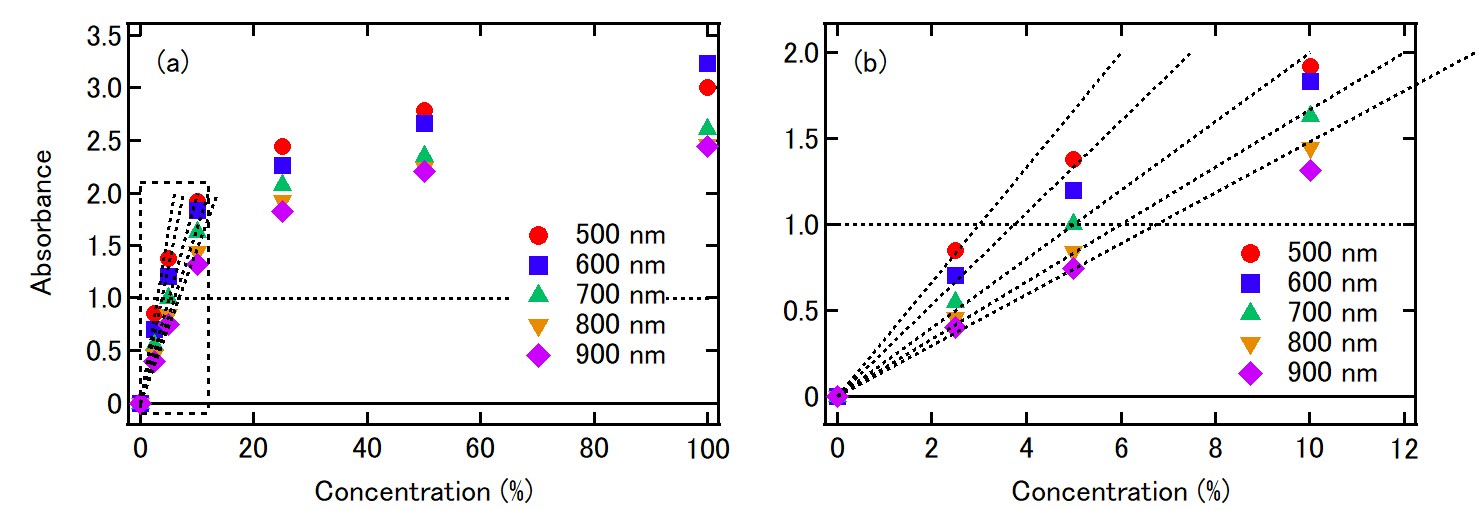}
\caption{(a) Absorbance versus concentration plots for tomato juice at five wavelengths (500, 600, 700, 800, and 900 nm) over the full concentration range. (b) Expanded view of the dilute region (0-12\% v/v), emphasizing the linearity predicted by the Beer-Lambert law.}
\label{fig:beerlaw}
\end{figure}

From an educational perspective, the sequence of Figures 2-4 allows students to connect qualitative observation (color change) with quantitative evidence (spectral data and fitted plots). This progression reinforces the scientific process of moving from observation to analysis and interpretation. The experiment encourages critical thinking as students identify possible causes of deviation and propose experimental improvements, such as filtering particulates, changing path length, or testing different wavelengths. Because the setup is easily modified, instructors can extend the activity to inquiry-based projects that compare different natural pigments or use LED sources to study wavelength dependence. Thus, the tomato-juice experiment not only demonstrates a core optical principle but also promotes creativity, teamwork, and reflective discussion about the limits of physical laws.

\section{Conclusion}
This work presents a simple and engaging experiment for demonstrating the Beer-Lambert law using diluted tomato juice and a halogen lamp. Students can directly observe how the intensity of transmitted light decreases with concentration and identify the characteristic absorption band of lycopene that gives tomatoes their red color. When plotting absorbance against concentration, they find linear behavior at low concentrations and clear deviations at higher ones, providing a natural opportunity to discuss the limits of ideal physical laws and the influence of scattering and turbidity.

From an educational perspective, the activity connects mathematical models of exponential attenuation with real optical measurements, enabling students to move from qualitative observation to quantitative analysis. The experiment reinforces scientific reasoning and encourages reflection on measurement uncertainty, data fitting, and the meaning of physical approximations. Because all materials and instruments are safe and portable, the setup can be implemented in a variety of educational settings, from classroom demonstrations to inquiry-based student projects. This approach thus transforms a familiar everyday material into a powerful tool for cultivating conceptual understanding and experimental curiosity in optics and physics education.


\ack{The author thanks the students who participated in the experiment for their enthusiasm and constructive feedback.}



\data{The raw data files are available at https://github.com/hwadati/tomatojuice.}

\suppdata{The Supporting Information is available at https://github.com/hwadati/tomatojuice.}


\end{document}